# NOSTOS – Spherical TPCs

**Abstract:** A new concept, the spherical TPC, presents unique advantages when low energy neutrinos are to be detected. Some of the applications are: a) observation of the $\nu_e$ oscillation when emitted by a tritium source ($<E\nu_e> = 14$ keV) in a 10 m radius sphere. This project, baptized NOSTOS, intends to give the mixing angle $\theta_{13}$, the Weinberg angle and a much lower limit ($< 10^{-12} \mu_B$) of the neutrino magnetic moment, b) an array of small (radius 2-4 m), cheap spheres scattered around the world would be perfect to monitor extragalactic supernovae, c) detect the spectrum of low energy (pp – $^7$Be) solar neutrinos.

P. Gorodetzky, for the NOSTOS collaboration. IN2P3 / CNRS, PCC – College de France, 11 place Marcelin Berthelot, 75005 Paris, France

**Introduction:** When detecting low energy neutrinos (0 to 1 MeV), one is faced with the huge problem of radioactive background. Apart from some unlucky ideas on inverse ß-decay experiments, the detecting medium is free from background. This one arises only in the medium envelope. The geometry maximizing the volume relative to the surface is the sphere. Furthermore, if the medium is a gas under pressure (generally needed to detect neutrinos), the sphere is the geometry that, to hold a given pressure, will have the thinnest envelope, hence be the less radioactive.

In the different detecting methods, the Time Projection Chamber (TPC) is the most convenient for its high accuracy in yielding the 3D coordinates of the event and its high rate possibility.

That is why the spherical TPC idea was born and is developed through 4 projects.

**Tritium Sphere:** The compilation of the experimental results of different neutrino detectors: solar, atmospheric or nuclear reactors have the oscillation with the following parameters: The atmospheric neutrino oscillation data [1, 2] strongly suggest that oscillate into with maximal mixing angle ($\theta_{atm} = 2$) and a corresponding mass squared difference of $\Delta m_{23}^2 \approx 3 \cdot 10^{-3} eV^2$.
$\delta m^2_{12} = 6 \times 10^{-5}$ and $\delta m^2_{13} = 2.5 \times 10^{-3}$

The other mixing angles being already determined, the challenge today is to measure the mixing angle $\theta_{13}$ which has a mall value ($\sin^2 2\theta < 0.2$), hence more difficult to determine, but will open the way to CP violation studies in the neutrino sector.

The idea is to find a source of low energy neutrinos. Tritium provides just that as fig.1 shows. The peak of the neutrino energy spectrum is 14 keV if the threshold for the electron (electron issued from elastic scattering).

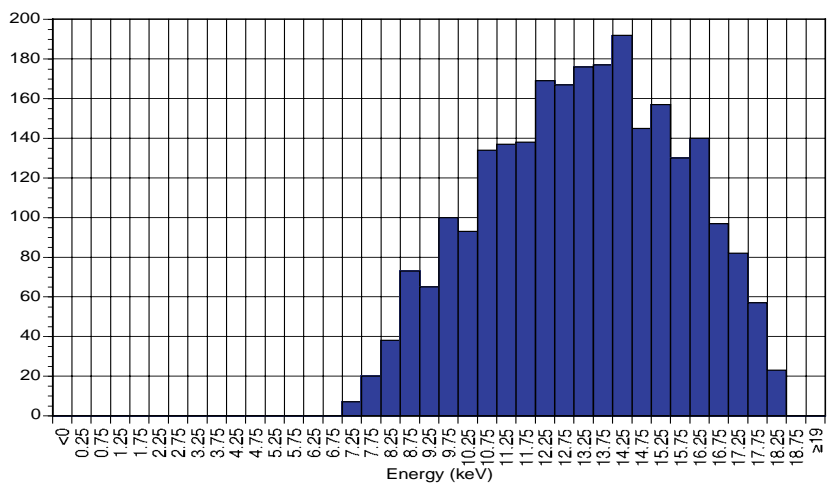

Fig.1 Neutrino energy spectrum of $^3$H

For that energy, the oscillation length (half wave) is just 10 m. A spherical TPC with the tritium in a central sphere of about 1 m diameter, a drift for the elastically scattered electrons from the external sphere (20 m in diameter) and the ability to detect the electron blob when it arrives at the central sphere is all it takes to do the job. If there are 20 kg of tritium (available from Russia or canadian electric power companies) and an optimistic 10 bar of Xe, the rate will be about 10000 events per year. The electric field at a radius r between two concentric spheres of radii $R_1$ and $R_2$, with a voltage $V_0$ applied to the central sphere (the external one being at ground), is

$$E = (V_0/r^2) R_1 R_2 / (R_2-R_1) \text{ in Volts / cm}$$

The time dispersion can be computed as being $\sigma_t \approx E^{-3/2} \approx r^3$ which shows another advantage of the spherical TPC: the determination of impact radius from the blob size is about 10 cm.

The central sphere can be set to a given voltage (here 500 kV to 1 MV) either by holding it with a tapered rod, or, better with a charging system similar to the Pelletron (chain of small metallic spheres linked by an insulator) used in some electrostatic accelerators. The power inside the central sphere can be taken from the movement of the "Pelletron", (or a small insulator rod…). The heat (some kW) generated by the tritium has to be evacuated in an insulating way.

The electron collectors will be Micromegas gas detectors, positioned all around the central sphere. Signals can be sent to the external world optically. The threshold should be very low, and some 200 eV are reachable (100 eV has already been reached in the Modane tunnel). The gain has to be high, $> 10^4$, and the HELLAZ R&D has shown this is possible with a good choice of quencher gas and the use of multi layered Micromegas.

Fig. 2 shows the oscillation simulated for 10 bar Xe in 4 months. Here $\sin^2 2\theta < 0.17$ has been taken.

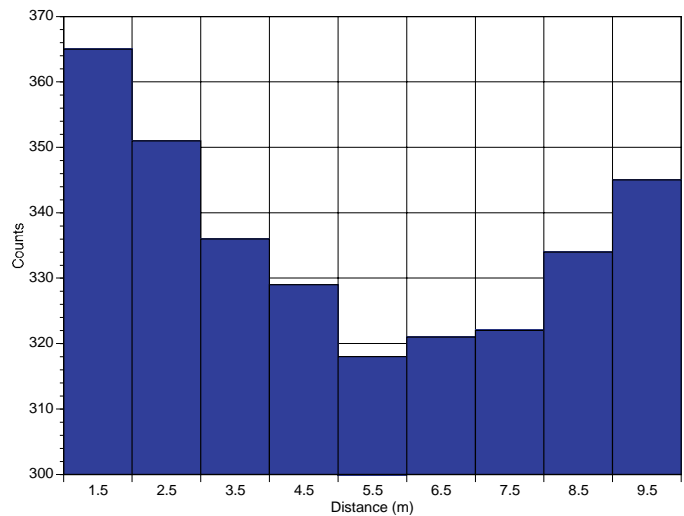

Fig. 2 Counting rate (300 bottom, 370 top of scale) versus radius (1 to 10 m).

In case the gain is too difficult to obtain, operation at 1 bar would require 3 years for the same results.

Neutrino magnetic moment: VOIR TEXTE1
Weinberg angle: VOIR TEXTE2

# Supernovae detection – neutrino coherent scattering  TEXTE3

**Low energy (pp – $^7$Be) solar neutrinos spectrum:** The determination of the low energy spectrum of the solar neutrinos is of utmost interest. Up to now, the only experiments which have the possibility to extract a spectrum are the water Cerenkov type: SuperK and SNO. However their energy threshold is high enough they see only $^8$B neutrinos. The models for the 8B are much less precise than those for pp or 7Be. The only solution to see neutrinos with an energy as low as 200 keV is the use of high pressure TPC. The HELLAZ project, uses 20 bar He where the lowest energy scattered electron (100 keV) has a range of some 5 cm, out of which the first 2 can be used to reconstruct its original direction. As the electron energy and the sun position is known, the neutrino energy can be reconstructed. The angular resolution on the scattered electron has to be of the order of 2° to yield a 10% neutrino energy resolution. In He, 40 eV are necessary to ionize the gas. Hence, 100 keV electron will make only 1000 ionisation electrons in the first 2 cm. To reach the necessary angular resolution, it is mandatory to use all possible spatial information, that is to know the xyz coordinates of each of the 1000 electrons. The detector at the end of the TPC must have a single electron recognition, that is a gain of the order of $10^6$. HELLAZ is an ongoing R&D project in Saclay. The gain at 20 bar He is now obtained by using double stage Micromegas (this stabilizes the electric field of the gap). X-Y readout was a problem: matrixing strips was nearly impossoible with the double gap. Now, pixels are used and the result seems promising. The original project HELLAZ was based on a usual cylindrical TPC. However, the radioactive backgroung was evaluated to 10000 events per day that are rejected hopefully by the knowledge of the direction (simulations say so). If we use a spherical TPC, the background can be reduced to 1/3. This is a great step forward. The sphere would be similar to the tritium sphere (10 m radius, to get about 20 solar neutrinos per day at 20 bar.) To get the coordinates of the ionization electrons, the micromegas pixels have to be about 3 mm in size, without magnification. This means that the TPC anode where the ionization electrons are collected cannot be the internal small sphere (as in the above projects) but the external one. This actually makes the readout Micromegas readout more easy.

**Prototype:** VOIR TEXTE4 et a la fin, mettre:
Improvement of the electrostatic issues: two methods are being evaluated. One is to have a tapered rod to hold the central sphere with conductive rings at the correct potential to respect the spherical field. In the last centimeter next to the center, the rod would be ended by a cone holding the small sphere on its summit. This cone, made of resistive material in its volume would very well respect the electrostatics. The second method is to charge the central sphere by injecting charges onto it. In a way similar to electrostatic accelerators, a chain of two mm metallic spheres, isolated one from each other by a nylon wire, would circulate between the external and internal spheres. Charging at the ground would be assured by a high voltage supply. Contrary to an accelerator, no beam discharges the center. This is why the chain can be very small.

Conclusions: The spherical TPC is a very promising. It is simple, does not require a field cage, has about 1/3 of the material necessary to hold a given volume. Associated with Micromegas, it will be a very powerful tool in the low energy neutrino study.